\documentstyle[twocolumn,aps]{revtex}

\input{epsf}

\input{epsf}

\begin{document}



\title{SCALING OF PARTICLE PRODUCTION WITH NUMBER OF PARTICIPANTS
IN HIGH-ENERGY A+A COLLISIONS IN THE PARTON-CASCADE MODEL}

\author{Dinesh Kumar Srivastava$^1$ and Klaus Geiger$^2$}

\address{$^1$ Variable Energy Cyclotron Centre, 1/AF Bidhan Nagar, 
Calcutta 700 064, India}

\address{$^2$ Physics Department, Brookhaven National Laboratory, Upton,
 N. Y. 11973, U. S. A.}

\date{\today}

\maketitle

\begin{abstract}

In view of the recent WA98 data of $\pi^0$ spectra from 
central $Pb+Pb$ collisions at the CERN SPS, we analyze
the production of neutral pions for
$A+A$ collisions  across the periodic table at 
 $\sqrt{s}=17$ A GeV and 200 A GeV within the framework of
 the parton-cascade  model for relativistic
heavy ion collisions. The multiplicity of the pions (having $p_\perp > 0.5$ GeV/c)
in the central rapidity region, is
seen to scale as $\sim (N_{part})^{\alpha}$,
where $N_{part}$ is the number of participating nucleons, which we 
have approximated as $2A$ for central collisions of identical
nuclei.  We argue that the deviation of $\alpha$ ($\simeq 1.2$) from unity
 may have its origin in the multiple scattering suffered by the partons. 
We also find that the constant of proportionality in the above
scaling relation increases substantially in going from
SPS to RHIC energies.  This would imply that the (semi)hard partonic 
activity becomes a much cleaner signal above
the soft particle production at the higher energy of RHIC, and thus 
much less dependent on the (lack of) 
understanding of the underlying soft physics background. 
\end{abstract}

\vspace{1.0cm}

Recently, the WA98 Collaboration has published \cite{wa98}
 data for the production of neutral pions up to
transverse momenta of $p_\perp \simeq 4$ GeV/c,
in central $Pb+Pb$ collisions at 160 A GeV/c incident momentum,
corresponding to $\sqrt{s}\simeq 17$ A GeV.
Two most interesting features of these data emerge when
compared to corresponding data from $pp$ collisions and
collisions involving lighter nuclei \cite{wa98}:
a) an approximate invariance of the spectral shapes, i.e.,
a near indepence of the slope of the neutral pion $p_\perp$ spectra;
b) a simple scaling of the $\pi^0$ with the number
of participating nucleons, if the number of participants is large
($ \,\lower3pt\hbox{$\buildrel > \over\sim$}\,30$).
\medskip

Soon after its publication the  (preliminary) WA98 data~\cite{prel}, 
was analyzed in two interesting papers, one by Wang \cite{wang} and 
the other by Gyulassy and Levai \cite{miklos}, 
with regard to the contribution to $\pi^0$ production at larger
 $p_\perp \,\lower3pt\hbox{$\buildrel > \over\sim$}\,2$ GeV/c 
due to (semi)hard parton scatterings and associated minijet production.
Astonishingly,
Wang was able to reproduce the WA98 data for the invariant $\pi^0$
cross-section by simply using the  perturbative QCD  cross-section
for a single parton-parton scattering
convoluted with the nuclear overlap function
and a quark/hadron fragmentation function,
and a parametrized accounting of the $p_T$-kick suffered by the
partons.
On the other hand, Gyulassy and Levai 
made a detailed numerical simulation using the event generator HIJING, and  
claimed that the consideration of multiple scattering effects of partons 
are crucial and not at all in conflict with the WA98 data.
\medskip

In the present paper, we analyze the WA98 data from yet
 another angle \cite{remark1}.
We would like to demonstrate that the observed scaling of the pion yield
in $A+A$ collisions as well as the insensitivity of the $p_\perp$ 
slope to a variation of $A$,
{\it does not at all} imply the absence of dense matter effects on the
particles, {\it nor} does it necessarily lead to the 
conclusion of Ref. \cite{wa98}
that the produced matter is thermally equilibrated.
Gyulassy and Levai have already hinted this, but our findings go even further.
In fact, a significant partonic cascade activity which scales roughly
with $A^{4/3}$ (rather than with $A$) does {\it not} lead to a contradiction
with the WA98 data,
rather, we believe that it is confirmed by the  WA98 results. 
\smallskip

In order to elucidate this apparent
 mystery, we use the event generator VNI \cite{vni} which embodies the physics
of the parton-cascade model~\cite{pcm} for 
ultra-relativistic heavy-ion collisions.
The model attempts to describe the nuclear dynamics
on the microscopic level of particle transport and interactions,
by evolving the multi-particle system in space-time from the instant
of nuclear overlap all the way to the final-state hadron yield. 
For details we refer the interested reader to Refs. \cite{vni}. Here 
it suffices to summarize the essential elements embodied in the model, namely: 
(i) the initial minijet production through liberation of partons 
from the colliding nuclei
by means of (semi)hard parton scatterings; 
(ii) the subsequent parton cascading consisting of multiple gluon emission and 
successive rescatterings;
(iii) the coalescence of final-state partons to color-neutral pre-hadronic
 clusters;
(iv) the decay of clusters into primary hadrons; 
(v) the subsequent hadron cascading consisting of reinteractions of
 hadrons and clusters with resonance production and decay.
\medskip

In this model, `dilute' collision systems (involving beams of protons or
light nuclei) naturally show a very different dynamical evolution than
`dense' collision systems (head-on collisions of heavy nuclei).
Whereas in the former case multiple particle interactions are absent or
negligible, in the case of heavy ions, 
the space-time development is characterized by multiple scatterings of 
partons, and later also of hadrons which are formed when the
partonic system hadronizes. 
Moreover, the parton cascading during the
early stage has rather distinct features, as compared to the
hadron cascading during the late stage of the collisions.
Multiple interactions among produced hadrons 
mostly alter only the momentum distributions \cite{ron},
and do not lead to  a substantial increase of particle production. 
The interactions among partons, on the other hand, have
a very different effect on particle production.
Perturbative QCD tells us that primary  (semi)hard scatterings of partons with
momentum transfer 
$q_\perp \,\lower3pt\hbox{$\buildrel > \over\sim$}\,1-2$ GeV/c
contribute the bulk of minijets. Moreover, the so kicked-out partons
are excited off-shell by an amount $Q^2 \simeq q_\perp^2$, which they
tend to shake off by gluon brems-strahlung.
Thus, even in the absence of partonic rescatterings, the additional emission 
of gluons  leads to an increase of the 
parton multiplicity on top of the number of those quarks and gluons
which are liberated by the primary scatterings.
In the parton-cascade model the characteristics of parton multiplication 
is further pronounced due to secondary interactions: 
Firstly, the initially kicked-out primary partons may rescatter
and receive additional momentum transfer that again feeds the gluon emission.
Secondly, all the newly produced off-spring of secondary partons 
increase the local density and can themselves re-interact.
However, upon hadronization, the partonic color charges have to reorganize
themselves to color-singlet composite objects that are the seeds of
the emerging hadrons. Clearly, in order to make a hadron of mass $m_h$,
the total invariant mass of the recombined partons must be at least
equal to $m_h$, and consequently it requires in the mean more than just
two partons to satisfy this kinematic condition, especially
since the bulk of gluons from radiative emission piles up
at the low-energy end of the perturbative regime.
Nevertheless one would expect a proportionality between
produced partons and resulting hadrons.
\medskip

Indeed,
the cluster-hadronization scheme \cite{EG}, employed here 
to convert the outcome of the parton cascade into hadronic states,
provides that the
number of produced  pre-hadronic clusters, and hence hadrons is proportional
to the number of partons present in the system. Thus, the production 
of particles should reflect the extent of partonic scattering 
throughout the nuclear collision.
This can be a very important consideration as the number of collisions
among the primary partons has been shown to scale linearly with the
number of the participating nucleons in the nuclear collision~\cite{sg_mult}.
Thus a deviation of the number of produced hadrons from a similar linear scaling
would be a direct confirmation of multiple scatterings taking place 
in the wake of relativistic collision of nuclei.
\medskip

A simple consideration may illustrate these features
of particle production. Let $x$ denote the number of
partons in each nucleon, and let each parton suffer $\nu$ collisions
during the partonic stage. Assuming that each virtual parton radiates
$r$ partons, we see that the number of produced partons will
vary as
\begin{equation}
N_{\mbox{\footnotesize{partons}}}\,\;\propto\,\; \nu \; (1+r) \;\, x \;\, A 
\;,
\end{equation}
and if, as proclaimed,
\begin{equation}
N_{\mbox{\footnotesize{hadrons}}}\,\;\propto\,\;
   N_{\mbox{\footnotesize{partons}}}
\;,
\end{equation}
one realizes immediately that if the partons interact only once, 
the multiplicity of the partons, and hence the multiplicity of hadrons, 
will scale as $A$. It is also clear that if every parton 
interacts with with every other
parton then $\nu \propto A$, and the number of materialized partons 
would scale as $A^2$. That can happen, if the system would live for an 
infinitely long time.
However, this is not the case. Rather than that, in 
relativistic heavy ion collisions, the partonic matter will expand,
dilute, and eventually convert into hadrons. Thus a given parton may
undergo $\nu \sim R/ \lambda$ 
interactions; where $R$ is the transverse 
size of the system and $\lambda$ is the mean free path of the parton.
Noting that $R\sim A^{1/3}$,  we immediately see that the
number of materialized partons, and hence the number of produced particles
would scale as $\approx \, A^{4/3}$. An experimental verification of this
scaling behaviour could be a direct manifestation the formation
of a  dense partonic matter!
\medskip

We shall demonstrate now that these simple considerations are 
indeed confirmed by a detailed simulation
with the event generator VNI on the basis of the 
parton-cascade/cluster-hadronization model.
We first consider the recently measured transverse momentum
distribution of $\pi^0$-production in central collsions of $Pb+Pb$
at CERN SPS obtained by the WA98 collaboration~\cite{wa98}. 
To make contact with the experimental data, 
the simulations were done
for the range of impact parameters $0 < b < 4.5 $ fm, which corresponds to 10\% 
of minimum-bias cross-section. The result of our  model calculation,
shown as the solid histogram in Fig. 1a, is seen to be in decent agreement with 
the experimental measurements.
The model results do not include the final-state interaction
among produced hadrons yet, but it is likely \cite{ron} that the agreement will 
further improve once the effect of cascading hadrons is included. 
The dashed histogram in Fig. 1a corresponds to 
$pp$ collisions at $\sqrt{s} =$ 17 GeV, and scaled accordingly. Comparing
$Pb+Pb$ with $pp$, clearly exhibits
the well-known enhancement of the production of pions having large
$p_\perp$ due to multiple scatterings among the partons~\cite{sg_mult}.
\medskip

As mentioned already,
one of the interesting observations of the WA98 results is 
near independence of the slope of the $p_\perp$ spectra for the pions once
the number of participants is large.
In Fig. 1b we plot our results for the $p_\perp$ spectra of $\pi^0$'s for 
a number of central $AA$ collisions at $\sqrt{s}=$ 17 A GeV for
various $A+A$ systems from $A=16$ to  $A= 197$. One observes that
they are almost identical in shape with a universal slope for 
$p_\perp \,\lower3pt\hbox{$\buildrel < \over\sim$}\,1.5$ GeV/c.
On  the other hand, the deviations appearing at larger $p_\perp$ for heavier
systems are indicative of enhanced multiple scattering there.
Similar results (not displayed here) were obtained at RHIC energies.
\medskip

\setcounter{figure}{0}
\begin{figure}[htb]
\epsfxsize=230pt
\centerline{ \epsfbox{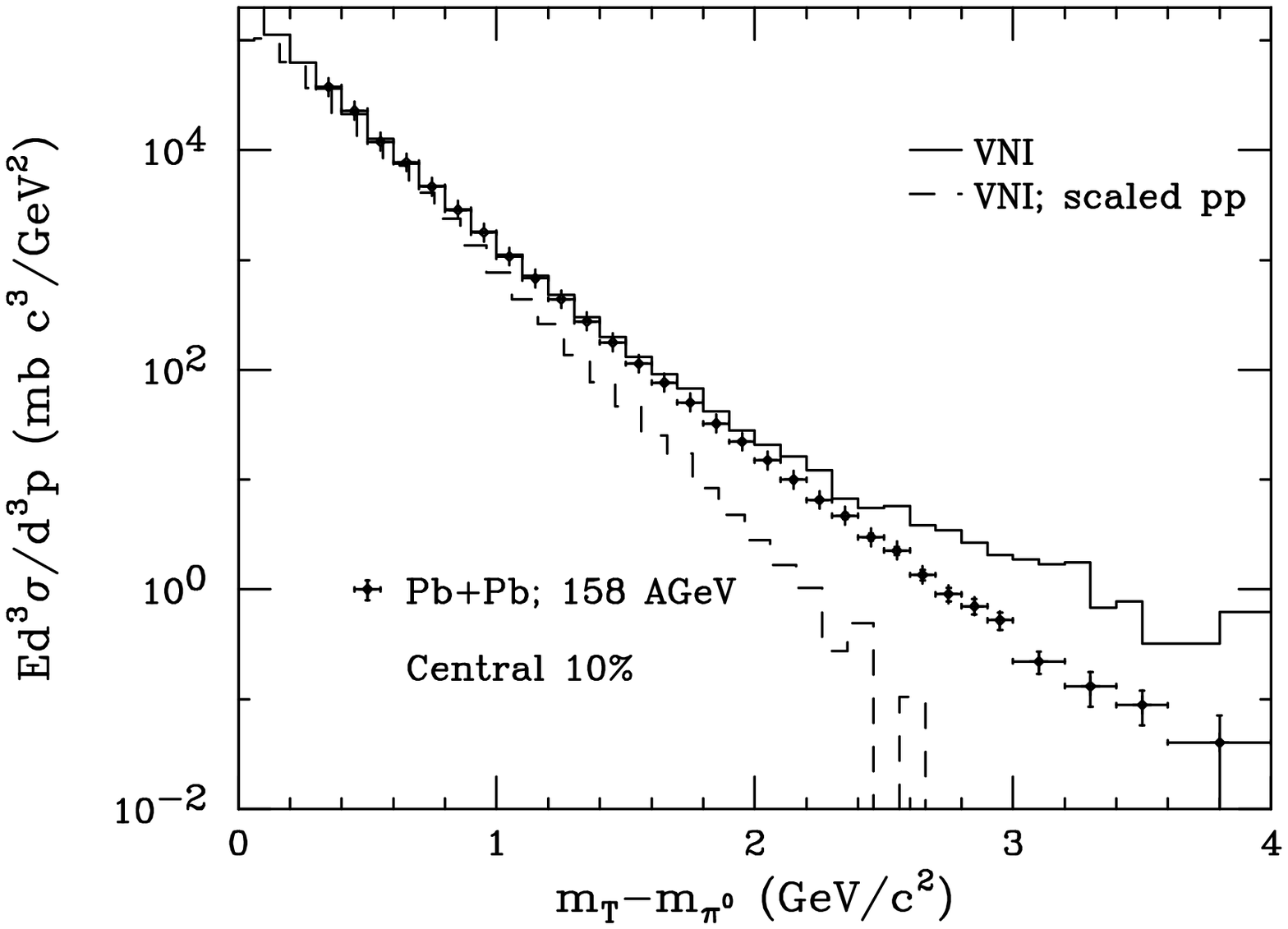} }
\end{figure}
\begin{figure}[htb]
\epsfxsize=230pt
\centerline{ \epsfbox{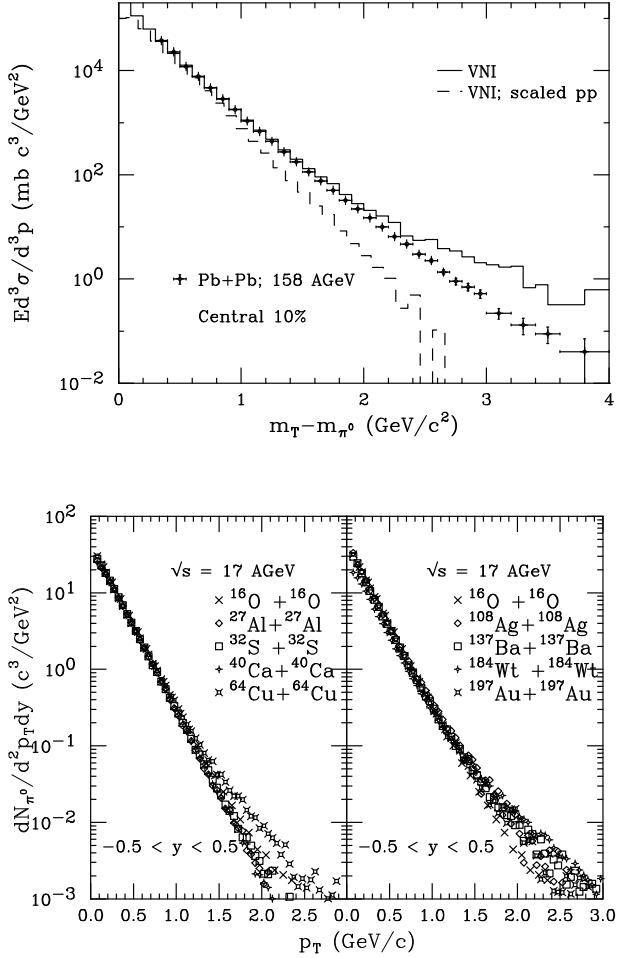}  }
\vspace{0.5cm}
\caption{
{\it a) Top}:
Transverse mass spectra of neutral pions in central (10\% of minimum
bias) collision of 158 AGeV $Pb+Pb$ collisions. The solid histogram 
represents our result
from the parton-cascade/cluster-hadronization model. The dashed histogram
corresponds to  $pp$ collsions at $E_{cm}=17$ GeV, scaled accordingly.
{\it b) Bottom}:
Transverse momentum spectra of neutral pions in central
collisions of identical nuclei at $E_{cm} = 17$ A GeV.
The symbols are our results for various systems, from $O+O$ up to $Au+Au$.
All results  are normalized to the case  $A=16$.
}
\end{figure}

In order to verify 
this scaling more closely, we have calculated, as a function of the nuclear
mass number $A$, the production of $\pi^0$'s
in the central rapidity region ($-0.5 < y < 0.5$)
having transverse momenta $p_\perp \ge 0.5$ GeV/c. The latter choice minimizes
the influence of pions having their origin in decay of resonances. This
kinematic window was motivated \cite{wa98} by the WA98 collaboration
in their measurement of the $\pi^0$ yield.
Fig. 2a displays the simulation results for central $A+A$
collisions across the periodic table, at CERN SPS center-of-mass energy
$\sqrt{s} = 17 $ A GeV, while Fig. 2b shows the same for RHIC energy 
$\sqrt{s} = 200 $ A GeV.  The solid lines are  fits to the model results,
represented by the symbols, and scale as 
\begin{equation}
N_{\pi^0} \;\propto\; \left( N_{part}\right)^{\alpha} 
\label{prop}
\;,
\end{equation}
where $N_{part}=2A$ is the number of participating nucleons, and
$\alpha$ being extracted as:
\begin{equation}
\alpha
\; \approx \;
\left\{
\begin{array}{l}
1.16 \;\;\; \mbox{at}\;\;\sqrt{s}=17 \;\mbox{A GeV} \\
1.23 \;\;\; \mbox{at}\;\;\sqrt{s}=200 \;\mbox{A GeV} 
\label{alpha}
\end{array}
\right.
\label{prop1}
\;.
\end{equation}
In view of the uncertainties, we may say that
$\alpha \approx 1.2$ is a fair number that characterizes
our model results.
This is quite similar \cite{remark2} to the observation of the WA98 
experiment at 
$\sqrt{s}\simeq 17 $ A GeV, namely \cite{wa98} $\alpha \approx 1.3$,
with the number of participating nucleons estimated from the
impact parameter, in collisions involving lead nuclei.
\medskip

\setcounter{figure}{1}
\begin{figure}[htb]
\epsfxsize=240pt
\centerline{ \epsfbox{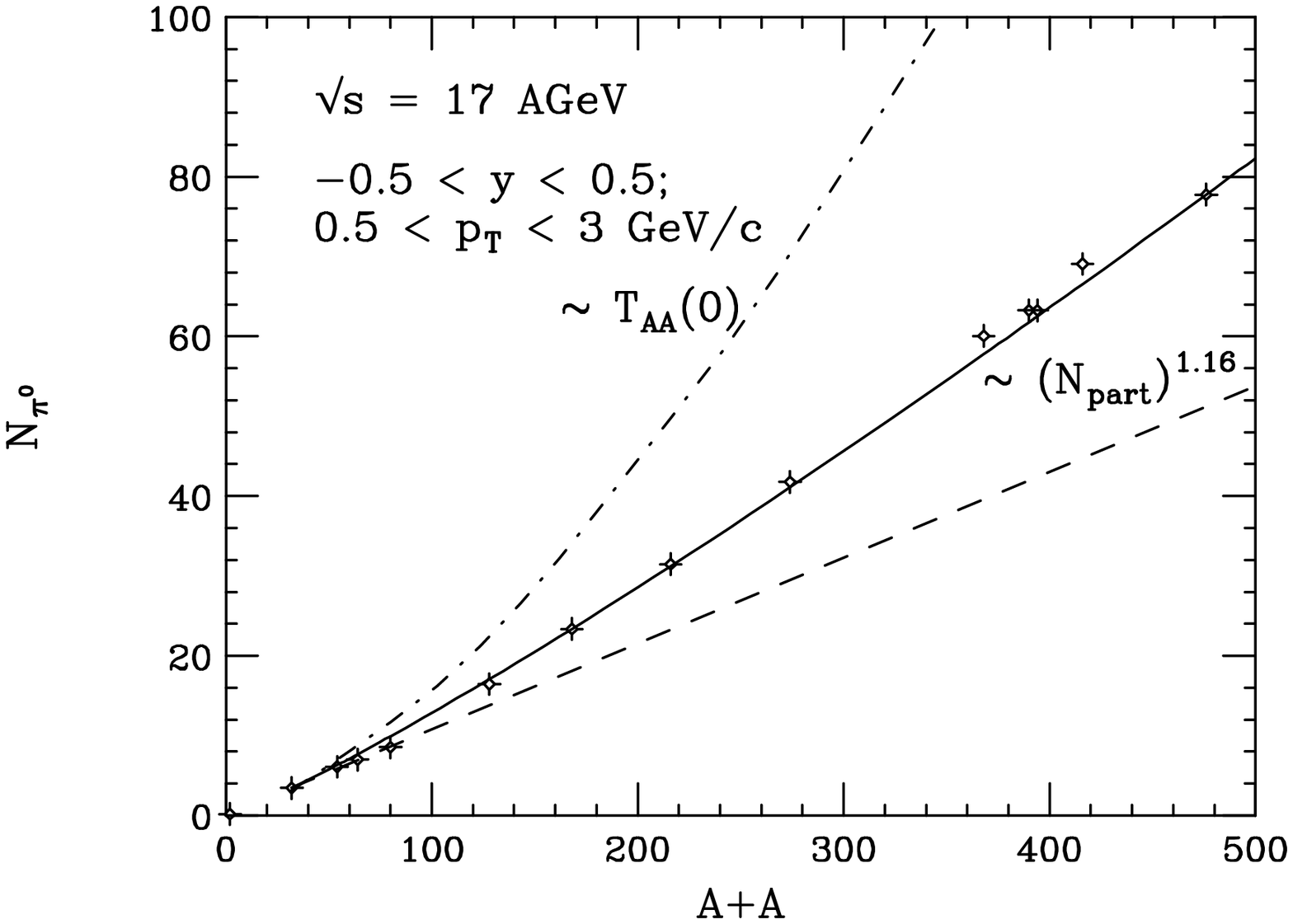} }
\end{figure}
\begin{figure}[htb]
\epsfxsize=240pt
\centerline{ \epsfbox{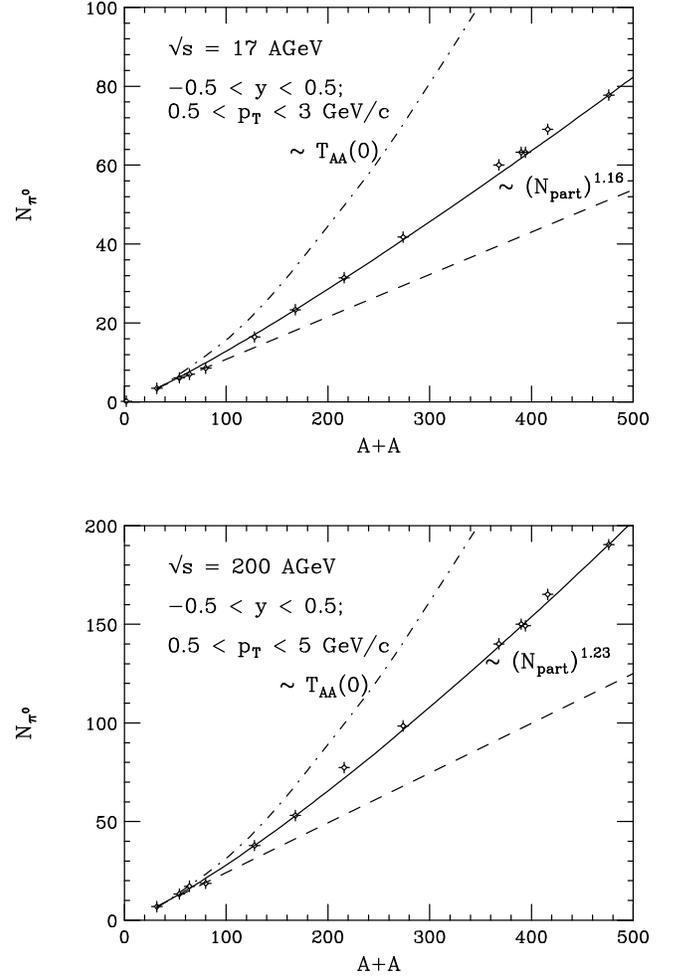}  }
\vspace{1.0cm}
\caption{
{\it a) Top}:
Mass number scaling of pion production in the central rapidity region
at CERN SPS energies.
 The symbols represent the results of
our simulations, and the solid curve is a fit to these results.
{\it b) Bottom}:
Mass number scaling of particle production in the central rapidity region
at  BNL RHIC energies. The symbols represent the results of
our simulations, and the solid curve is a fit to these results.
Dashed lines would correspond to a linear scaling, dotted-dashed lines
to a scaling with the nuclear overlap $T_{AA}(b=0)$.
}
\end{figure}
\medskip

For comparison, the dashed lines correspond to a linear scaling
$ \sim ( N_{part})^{1.0}$, whereas the dashed-dotted
lines indicate a hypothetical scaling with the nuclear overlap factor
$T_{AA}(b=0)\sim (N_{part})^{1.42}$. 
A linear scaling would reflect a single-collision situation,
and a scaling with $T_{AA}$ would indicate a Glauber-type multiple-collision 
scenario, in which nucleons suffer several collisions
along their incident straight-line trajectory, without deflection but
with energy loss.
Comparing the three curves, we can conclude that
our simulation results rise significantly slower than
with $T_{AA}$, because firstly, the particles change direction through the
collisions, and secondly, they are subject to a collision time of the order
of the inverse momentum transfer, during which they cannot rescatter.
On the other hand, our calculated $\pi^0$ yields grow much faster than
linear with $A$, due to multiple scatterings.
Hence, the scaling of the $\pi^0$ yield in the model may be interpreted
as a {\it collision-meter} that indicates the extent of multiple interactions
on the parton level (recall, that we did not include hadronic final-state
interactions). 
Such an interpretation would also explain the (slightly)
larger value of $\alpha$, and also the constant of proportionality 
(cf. Fig. 2a and 2b)
i.e. the stronger increase, at RHIC energy 
as compared to the results for CERN SPS energy.
In the former case one would expect that multiple collisions are
further enhanced due to a higher density of produced particles.
This, we find, results in a increase larger than a factor of 2
in the number of $\pi^0$'s between CERN SPS and RHIC. 
\bigskip

In summary, we have demonstrated here that the observed scaling with 
$A^{\alpha}$
with $\alpha\simeq 1.3$
of the number of produced particles in heavy-ion $A+A$ collisions, as well as
the approximate shape-independence of the transverse momentum spectra, 
are satisfactorily reproduced by the parton-cascade / cluster-hadronization
 model, though with a slightly smaller value of $\alpha\simeq 1.2$.
 The  quark-gluon multiplicity from the early stage of
the collisions due to initial minijet production {\it plus}
subsequent cascading with multiple scatterings and gluon emissions, leads to a
 final-state particle multiplicity which approximately scales with the number
 of nucleon collisions (i.e.,  $\sim A^{4/3}$), as inferred
 by the WA98 experiment using high $p_\perp$ $\pi^0$'s. 

 The {\it key result} of our model simulations
 is that multiple parton  scatterings contribute significantly
 in collisions involving heavy nuclei.
 We have checked that
 the initial minijet production component alone, without
 subsequent cascading and reinteractions of the minijets,
 would yield for heavy ions a $\pi^0$ yield as for the $pp$ case in Fig. 1a,
 which  would clearly underestimate the WA98 $Pb+Pb$ data.
\smallskip 


\bigskip


One of us (DKS) would like to thank Terry Awes and Bikash Sinha 
for valuable discussions.
This work was supported in part by the D.O.E. under contract no.
DE-AC02-98CH10886.



\begin{thebibliography}{39}
\bibitem{remark2}
The  difference between $\alpha \simeq 1.3$ of the WA98 measurement,
and the  value $\alpha\simeq 1.2$ extracted from our calculations,
 may be attributed
to the specific way  particle collisions are simulated in VNI:
for each collision a characteristic collision time 
$\tau_{coll} = const./\sqrt{q^2}$
is estimated, depending on the invariant squared momentum transfer $q^2$. The
prefactor is not known, so we take $ const. = 1$, but one may consider
this as a true parameter to be adjusted.
 We also note that the geometry of central $A+A$ collisions is 
 with $2A$ participating nucleons is some what different from the
 case when $A$ nucleons from a Pb projectile-nucleus interact with
 $A$ nucleons from a Pb target-nucleus for some impact parmeter.
\end{thebibliography}
\end{document}